\documentclass[prb,aps,10pt]{revtex4}

\begin{document}
\title{Transport properties in manganite thin films}

\author{S. Mercone$^1$, C.A. Perroni$^2$, V. Cataudella$^2$,
C. Adamo$^1$, M. Angeloni$^3$, C. Aruta$^1$, G. De Filippis$^2$,}
\affiliation{\vspace*{-.7cm}}
\author{ F. Miletto$^2$,  A. Oropallo$^2$, P. Perna$^4$, A. Yu. Petrov$^1$, U. Scotti di Uccio$^4$, and
L. Maritato$^1$} \affiliation{\vspace*{.3cm}$^1$ Coherentia-INFM
and Dipartimento di Fisica ``E. R. Caianiello'', Universit\`a di
Salerno, Via S. Allende, 84081 Baronissi (SA), Italy \\$^2$
Coherentia-INFM and Dipartimento di Scienze Fisiche, Universit\`a
degli Studi di Napoli ``Federico II'', Via Cintia, 80126 Napoli,
Italy \\$^3$ Coherentia-INFM and Dipartimento di Ingegneria
Meccanica, Universit\'a di Roma Tor Vergata, Via del Politecnico
1, 00133 Roma , Italy \\\\$^4$ Coherentia-INFM and Di.M.S.A.T.,
Universita' di Cassino, Via Di Biasio 43, 03043 Cassino (Italy) }

\date{\today}

\begin{abstract}

The resistivity of thin $La_{0.7}A_{0.3}MnO_{3}$ films ($A=Ca, Sr$)
is investigated in a wide temperature range. The comparison of the
resistivities is made among films grown by different techniques and
on several substrates allowing to analyze samples with different
amounts of disorder. In the low-temperature nearly half-metallic
ferromagnetic state the prominent contribution to the resistivity
scales as $T^{\alpha}$ with $\alpha \simeq 2.5$ for intermediate
strengths of disorder supporting the theoretical proposal of single
magnon scattering in presence of minority spin states localized by
the disorder. For large values of disorder the low-temperature
behavior of the resistivity is well described by the law $T^{3}$
characteristic of anomalous single magnon scattering processes,
while in the regime of low disorder the $\alpha$ exponent tends to a
value near $2$. In the high temperature insulating paramagnetic
phase the resistivity shows the activated behavior characteristic of
polaronic carriers. Finally in the whole range of temperatures the
experimental data are found to be consistent with a phase separation
scenario also in films doped with strontium ($A=Sr$).

\end{abstract}

\maketitle

\newpage

\section{Introduction}

In the last years the mixed-valence perovskite manganese oxides
$La_{1-x}A_{x}MnO_{3}$ (where $A$=$Ca$, $Sr$) have been
intensively studied for their striking  properties such as the
colossal magnetoresistance ($CMR$). \cite{helm} The strong
sensitivity to the magnetic field is found in the range $0.2 < x <
0.5$ at temperatures $T$ around the ferromagnetic-paramagnetic
($FP$) transition point (the Curie temperature $T_C$) that is
often close to the temperature $T_p$ where a peak in the
resistivity signals the metal-insulator ($MI$) transition.
\cite{Salomon} The interplay between the $Mn$ magnetic moments
alignment and the metallic behavior is usually explained by
invoking the double-exchange ($DE$) interaction, \cite{Zener}
that, however, only qualitatively accounts for the properties
around the combined $FP$ and $MI$ transition, \cite{Millis}. As
shown by many experimental results, \cite{Salomon,Ibarra,Millis1}
other interactions, mainly the coupling of the charge carriers
with lattice, cooperate to drive the $MI$ transition and the $CMR$
effect. Actually a Jahn-Teller distortion of the oxygen octahedron
can lead to the trapping of the charge carriers into a polaronic
state influencing the transport properties in the high temperature
phase. In these compounds the $MI$ transition is affected by the
crystal structure also because of the dependence of the $Mn-Mn$
electron transfer matrix element on the $Mn-O-Mn$ bond angle whose
variation is a function of the radii of $La^{3+}$ and $A^{2+}$
cations. \cite{Hwang} Finally direct evidences of coexisting
insulating localized and metallic delocalized components have been
reported from many experimental techniques pointing out that the
tendency toward phase separation is intrinsic in these compounds.
\cite{dagotto} Indeed the phase coexistence arises from the
complex interplay between electron, orbital, spin and lattice
degrees of freedom affecting most properties of the system near
the phase boundary.

Even if a great effort has been done to understand the transport
properties of these materials, a complete comprehension of the low
temperature resistivity $\rho$ in the half-metallic ($HFM$)
ferromagnetic phase remains elusive. Indeed there is no agreement
on the dependence of $\rho$ as function of $T$ in this phase. The
law $\rho(T)-\rho_0 \sim T^2$, with $\rho_0$ residual resistivity,
has been proposed to fit the data of single crystals in the low
temperature range. \cite{Salomon,jaime,Urushibara,Okuda} For the
majority spin electrons the temperature dependence of the
resistivity due to the electron-electron scattering would provide
the $T^{2}$ dependence, however the $T^2$ term is about $60$ times
larger than the expected one for this type of scattering.
\cite{Salomon} Another source for this $T^2$ behavior would be the
single magnon scattering involving spin-flip processes,
\cite{mannari} but in a truly $HFM$ system this process is
suppressed since there is a band gap at the Fermi energy for one
of the spin channels. On the other hand, the two-magnon scattering
gives a $T^{9/2}$ dependence, \cite{kubo} that is in disagreement
with experimental data. Therefore in order to explain the behavior
of $\rho$, it has been argued that in single crystals at
intermediate temperatures the observed contribution could reflect
the reappearance of minority spin states that become accessible to
thermally excited magnons. \cite{Salomon,jaime} Of course this
single magnon process becomes possible only if the spin
polarization strongly decreases from unity with increasing $T$. In
any case, in single crystals some experiments have found
variations in the temperature scaling of $\rho$ from $T^2$ to
$T^3$ behavior, that is interpreted in terms of an anomalous
single magnon scattering process. \cite{Akimoto} This scattering
channel opens at finite temperatures where the $HFM$ structure of
conduction can break down and, as a consequence, the rigid band
approaches should not be justified. If one takes into account the
non rigid behavior due to spin fluctuations, the inverse lifetime
of the majority spin carriers is proportional to the density of
state of the minority carrier band as well as the magnon density
giving rise to the $T^{3}$ dependence in the low temperature
resistivity. \cite{Furukawa}

The situation in manganite films and ceramic samples is more
complicated. Some researchers have interpreted the temperature
behavior of the film resistivity as due to the $T^2$ term,
\cite{Snyder,Li} while others have attributed the low temperature
dependence of $\rho$ in $La_{1-x}Ca_{x}MnO_{3}$ and
$La_{1-x}Sr_{x}MnO_{3}$ films to the polaron coherent motion.
\cite{Zhao,Zhao2,Chen} Even if this latter process provides a good
fit of the resitivity, the model requires the existence of
exceedingly soft optical modes and polarons at almost zero
temperature. In $La_{1-x}Ca_{x}MnO_{3}$ systems the electrical
resistivity below $T_C$ has been fitted also by a $T^{2.5}$
dependence. \cite{Schiffer} This non conventional result has been
interpreted in these nearly $HFM$ compounds taking into account a
finite density of states of the minority spins at Fermi energy and
their Anderson localization. \cite{Wang} The spin-flip scattering
involving single magnons can occur with finite probability giving
a $T^{2.5}$ temperature dependence of the resistivity as result of
the exact solution of the linear response equation. Therefore also
in films and ceramics the transport properties at low $T$ are
considered to be strongly influenced by the single magnon
scattering. Finally, in contrast to the behavior of single
crystals, at high temperatures $La_{1-x}Sr_{x}MnO_{3}$ films are
characterized by a decrease of the resistivity with increasing $T$
signaling that an insulating phase becomes stable in the $CMR$
region. \cite{Snyder}

Recently the attention has focused on the role of the disorder in
these systems. \cite{dagotto} The effective strength of the
intrisic disorder is influenced by several quantities, such as the
tolerance factor since random potential fluctuations are due to
different sizes and electronegativities of $La^{3+}$ and $A^{2+}$
cations. The random disorder is important to smear the first-order
transition between competing states and to induce microscopic
inhomogeneities. \cite{Burgy} Besides, random potential effects
are able to give a large modification of the phase diagram near
the bicritical point between charge-orbital ordering and $FM$
states. \cite{Akahoshi} Actually the disorder suppresses the
charge ordering, shifting the phase boundary between the
ferromagnetic metal and the ordered insulator with respect to the
case of the clean compound. \cite{Motome} Even if the insulating
phase is not directly triggered by the disorder in many compounds,
\cite{Smolyaninova} the effect of randomness controls the value of
$T_c$ and the transport properties at least at low temperature.
\cite{Tiwari} Actually measurements in $La_{1-x}Sr_{x}MnO_{3}$
single crystals and polycrystalline compounds
\cite{Auslender,Zhang}, and $La_{1-x}Ca_{x}MnO_{3}$ films
\cite{Kumar,Ziese} have shown that the low temperature resistivity
exhibits a shallow minimum. There is a quite general consensus
upon the influence of electron-electron interaction with
scattering from static inhomogeneities as the dominant mechanisms
for the upturn. However it is not clear what is the role of the
disorder on the electrical resitivity for temperatures larger than
that of the minimum but smaller than $T_p$.

In this paper, we report on our measurements of resistivity in
$La_{0.7}Sr_{0.3}MnO_{3}$ ($LSMO$) and $La_{0.7}Ca_{0.3}MnO_{3}$
($LCMO$) films grown by different techniques and on several
substrates. The availability of samples prepared in  different
ways is of great advantage in this context for several reasons.
First of all, it is easier to address possible systematic errors
due to sample-dependent effects. In fact, the resistivity may be
determined not only by intrinsic mechanisms, but also by other
contributions such as grain boundaries, local defects and spurious
phases. Secondly, it is possible to study samples that differ for
strain and thickness, and that show different values of $T_p$ and
of resistivity. Since the residual resistivity is a measure of the
global disorder, this implies the possibility to investigate the
role of disorder in the transport properties of manganites.
Finally, the analysis has dealt with two classes of manganite
compounds ($LSMO$ and $LCMO$) since they are characterized by
different properties in the $CMR$ range. Indeed $LSMO$ systems
show the highest critical temperatures, weak-to-intermediate
electron-phonon ($el-ph$) coupling and disorder, whereas $LCMO$
systems demonstrate $MI$ transition at lower temperatures and
belong to the group of manganites with intermediate-to-strong
$el-ph$ and disorder strength. \cite{dagotto,perrofilm}

In section II of this paper we briefly describe the different
experimental techniques used for growing and characterizing films.
In section III we report the obtained results along with the
theory which supports them. Indeed in subsection III.A we carry
out a detailed study of the charge transport at low temperature
($T<T_{p}$). In the region of the ferromagnetic metallic ($FM$)
state the temperature contribution scales as $T^{\alpha}$ with
$\alpha$ close to $2.5$ for an intermediate range of residual
resistivities supporting the role of single magnon scattering
when, due to the disorder, the minority spin states are localized.
\cite{Wang} For large values of disorder the resistivity scales
with the law $T^{3}$ characteristic of anomalous single magnon
scattering processes, while in the regime of low disorder $\alpha$
tends to a value near $2$. These behaviors are quite robust since
they are independent of the film size and the strain distribution.
In subsection III.B the high temperature ($T>T_p$) resistivities
are discussed showing that the activated behavior characteristic
of polaronic carriers is present in the films. Finally in the
whole range of temperature the experimental data are consistent
with a phase separation scenario also in $LSMO$ films.

\section{Experimental}\label{due}

We considered $LSMO$ and $LCMO$ films prepared by different
techniques: a) pulsed laser deposition ($PLD$); b) molecular beam
epitaxy ($MBE$); c) Sputtering. We briefly summarize the
fabrication technique of the samples.

a) {\bf PLD}

The $PLD$ deposition was carried out by using a KrF excimer laser
($\lambda = 248$ $nm$) with a repetition rate of 3 Hz. The pulse
width was 25 ns, and pulse energy 150 mJ. The substrates have been
held at 700 $^{o}C$ in oxygen atmosphere ($P_{O_{2}}$ = 50 Pa) 50
mm far from the target. After film growth, the samples were cooled
at room temperature in about ten minutes in oxygen at 0.5 bar. Two
different samples have been considered in this work. The first one
is a 300 nm $LCMO$, the second a 160 nm thick $LSMO$. The first
sample has been deposited onto on a (100) $SrTiO_3$ ($STO$)
substrate, while the second on (100) $LaAlO_3$ ($LAO$). X-ray
diffraction ($XRD$) measurements in the Bragg Brentano
configuration yield a lattice spacing close to that of $LCMO$ and
$LSMO$ bulks, respectively. Considering the relatively high
thickness value, it is reasonable to assume that the stress due to
the substrate is completely relaxed in these samples.
\cite{Angeloni}

b) {\bf MBE}

Thin $LSMO$ samples on different substrates have been deposited by
$MBE$ in the same batch, using a co-deposition procedure in which
the elemental rates of $La$ (e-beam source), $Sr$ and $Mn$
(effusive cells) have been carefully controlled to obtain the
desired sample composition. The (100) $STO$, (110) $NdGaO_3$
($NGO$), and (100) $LAO$ substrates have been held at 700$^{o}C$
during growth. The peculiarity of the $MBE$ is the possibility to
achieve the in-situ formation of the perovskitic phase at very low
oxygen pressure without any post-annealing treatment. In this
case, a mixture of $O_2$ + 5\% Ozone at a total pressure P = $2.6
\cdot 10^{-2}$ Pa was employed. The atmosphere composition inside
the deposition chamber has been controlled by mass spectroscopy.
The reflected high energy electron diffraction ($RHEED$) analysis
has been performed during the growth process to check the
structural properties of the films. Through reflectivity
measurements we have studied the surface roughness and the
thickness of the thin films. Details of these surface analysis,
$EDS$, and X-Ray diffraction are reported elsewhere. \cite{Petrov}

c) {\bf Sputtering}

Several LSMO samples have been deposited by on axis RF magnetron
sputtering on various substrates, i.e., on(100) and (110) $STO$,
and on (110) $LAO$. The deposition conditions that give the best
samples in terms of cation stoichiometry, crystal structure, and
transport properties are the following. The sputtering pressure
(50\% $O_2$, 50\% Ar mixture) has been varied in the range 50 - 70
Pa. The substrates have been held at 840 $^{o}C$ 40 mm far from
the $LSMO$ target. Such samples are smooth, highly ordered, and as
a general rule present low resistivity and high Curie temperature.
Moreover, some films have also been deposited in non optimized
conditions, yielding samples with reduced $T_p$ and higher
resistivity. More details on the fabrication procedure and a
careful structural characterization of samples with both (100) and
(110) orientation are discussed in a separate paper.
\cite{Umberto} Briefly, the films deposited on both (100) and
(110) substrates grow in the usual cube-on-cube mode. The samples
deposited on (110) $STO$ ($SSS1$ in table \ref{samples}) are fully
strained with lattice parameters $a=3.89 \pm 0.01$ \AA, $b=3.89
\pm 0.01$ \AA, and $c=3.91 \pm 0.01$ \AA. The sample grown on
(110) $LAO$ is instead completely relaxed with lattice parameters
$a=b=c=3.89 \pm 0.01$ \AA.

\vspace{0.3cm} $XRD$ analysis in the Bragg Brentano configuration
has been performed on the produced samples after deposition in
order to characterize their crystal properties. For each technique
a very careful investigation of epitaxy, orientation, strain,
crystal quality and twinning was performed by using different
kinds of analysis. All the samples that have been analyzed in this
work show high structural quality. This has been assessed by the
observation of sharp rocking curves and of the interference
fringes around the reflections in $\theta -2 \theta$ scans.
Typical values of the rocking curve width are shown in Table
\ref{samples}. The stoichiometry of these samples has been
determined by energy dispersive spectroscopy ($EDS$) for samples
deposited on MgO with the same deposition parameters. Recently we
have also carried out Atomic Force Microscopy measurements which
confirm the low roughness and the good quality of the samples.
Some structural parameters of the representative samples
\cite{Angeloni,Petrov,Umberto} are summarized in Table
\ref{samples}. As reported in this table, there is a large variety
in the film thickness, strain and substrate orientation of the
films. All the characterizations allow us to confirm the
homogeneity of their growth and to affirm that the film are
epitaxial.

The measurement of the temperature dependence of resistivity in zero
magnetic field was performed in the standard four-probe
configuration, with the usual compensation of thermoelectric bias by
inversion of the direction of current flow. Electrical contacts to
the samples are provided by direct indium soldering on the manganite
film, or by soldering on Au pads deposited by sputtering. In both
cases, the contact area is $\sim 1$ $ mm^2$. The Van de Pauw
technique is employed to deduce the geometrical factor that allows
estimation of resistivity. \cite{Pauw} To this aim, thickness values
are provided by calibration to the oscillations of the X ray
reflectivity. We checked the error introduced by the geometrical
configuration of the contacts by repeated evaluations after removal
and replacement of the contacts. Also taking into account the
experimental error in the measurement of samples thickness, we
estimate that the error in the geometrical factor is about $ 10\%$.
Part of the measurements have been performed resorting to a
cryogenic inset in a He bath; in others a cryocooler was employed.
In both cases, we devoted a special care to the problem of sample
thermalization and temperature measurement. On the experimental
basis, both the measurement techniques lead to small spurious
thermal hysteresis in a cooling - heating cycle of measurement. The
effect on the determination of the physical parameters in the
fitting session is negligible, as discussed in the following.

\section{Results and Discussion}

The resistivity curves present general features. Indeed all the
samples are characterized by the $MI$ transition marked by the
temperature $T_p$. As shown in Table \ref{samples}, there is a
large variation in the values of $T_p$ according to the growth
technique, the thickness and the strain. Besides, below $20K$ the
resistivity of our samples shows a shallow upturn that has been
interpreted as due to quantum interference effects in presence of
disorder. \cite{Auslender,Zhang,Kumar,Ziese} Since this issue is
beyond the purpose of the present work, we have focused on the
temperature dependence of the resistivity starting from the
minimum.

In the following subsections the low and the high temperature
regimes of the resistivities will be studied in detail.

\subsection{Low temperature range}

The $\rho(T)$ plots of all the samples have been fitted by the
following function:
\begin{equation}
\rho(T)=\rho_{0}+AT^{\alpha}, \label{exponent}
\end{equation}
with $\rho_{0}$, $A$ and $\alpha$ free parameters. Here $\rho_{0}$
is the residual resistivity that we consider as a measure of the
effective disorder, and $AT^{\alpha}$ a generic T-power law which
can simulate different scattering processes. Typical values of the
residual resistivity $\rho_0$ are always less than $4*10^{-3}
\Omega*cm$ which can be considered a check of high quality of the
samples. Together with the results obtained on the behavior of the
$\rho(T)$ shown in the following, this feature represents an
important argument which supports the absence of any kind of
effect on the resistivity rising from grain diffusion. As observed
by Gupta \emph{et al.}, \cite{gupta} even grains of the order of
$10 \mu m$ have strong effects on both $\rho_0$ and $\rho(T)$ at
low temperatures. In fact, due to  the diffusion by grain surface,
the $\rho_0$ becomes higher and the behavior of $\rho$ changes as
a function of temperature. As it will be shown in the following
results, this is not the case for our resistivity data. Thus we
can conclude that grain boundary effect can be neglected in our
analysis.

In Fig. 1 we plot the resistivity measurements and the
corresponding fits of two representative samples (a $PSS0$ and a
$MSS0$) in three different ranges of temperature: (4 - 60)K, (4 -
120)K and (4 - 200)K. In Table II we report the parameters
$\rho_0$, $A$ and $\alpha$ defined in equation (\ref{exponent})
for the different fit sessions, together with the coefficient of
determination $R^2$. In any range of temperature, the fit provides
an excellent approximation of the experimental data ($R^2$ very
close to 1). The sensitivity of the fit to the value of the
parameter $\alpha$ has been checked in the following way. Once the
best fit parameters are determined, $\alpha$ is fixed at value
different from the optimal estimate, and $\rho_0$ and $A$ are
calculated by a new fit registering the variation of $R^2$. As a
rule, a variation $\Delta \alpha = 0.1$ leads to $\Delta R^2$
larger than $10^{-4}$. The comparison of the data in Tab.
\ref{fittable} also suggests the following considerations on the
reliability of the values of the fit parameters. First of all, it
is seen that $\rho_0$ is not affected by the choice of the fitting
interval. The statistical error on $\rho_0$ is negligible,
therefore the overall error is due to the experimental uncertainty
on the geometrical factor, as discussed in the previous section.
The case of $\alpha$ is different, because a variation is
typically observed when different ranges of temperature are
considered. Also the choice of the lower limit of the temperature
range deserves attention, because of the shallow upturn of
resistivity at low temperature. This region has been excluded from
the range of the fit by Eq. (\ref{exponent}), because they are out
of the limits of validity of the model. Our analysis of the data
leads to the conclusion that an overall uncertainty $\Delta \alpha
= \pm 0.1$ results from the different possible choices of the
temperature range. In view of the physical interpretation of this
parameter in the overall temperature range, we argue that this is
the uncertainty of the whole procedure (measurement and fit
session). Other experimental and statistical effects are in fact
negligible. As an instance, we checked that the error due to the
thermal coupling of the samples (i.e., the finite value of dT/dt
during the measurements, with consequent shift of temperature
between sample and thermometer) is well below $0.1$ in all
measurements.

In previous investigations the deviation from the quadratic power
law has been ascribed to a combination of terms due to different
kinds of scattering. \cite{Zhao, Li, Chen} In order to understand
if this analysis can be performed also for our samples, we have
carried out the fits of the data with some possible combinations
of terms. So we have used the following equation to fit the data:
\begin{equation}
\rho(T)=\rho_{0}+AT^{2}+\emph{S},
\label{eq2}
\end{equation}
where \emph{S} stands for the term due to the scattering with
anomalous single magnons \cite{Akimoto} ($T^3$), two magnons
\cite{kubo} ($T^{9/2}$), spin-waves \cite{Chen} ($T^{7/2}$),
acoustic phonons \cite{Li} ($T^{5}$) and optical phonons
(proportional to the phonon thermal distribution with the
frequency $\omega_0$ fit parameter). Even giving a larger weight
to the data at very low temperature, we did not succeed in
obtaining the excellent agreement that we obtained with equation
(\ref{exponent}) (it always provides the fit coefficient $R^2$
more close to 1). This results point out that the $T^{2.5}$
dependence in our samples cannot be simulated through a
combination of different power laws, as assumed previously.
\cite{Schiffer} Moreover this behavior finds a natural explanation
within a theory that considers the role of the disorder in nearly
$HFM$ systems. \cite{Wang} By taking into account a finite density
of states of the minority spins at Fermi energy and the Anderson
localization of them, the spin-flip scattering involving single
magnons can occur with finite probability. Resolving the linear
response equation, the temperature dependence of the resistivity
is given by $T^{2.5}$ starting from a low characteristic
temperature.

The analysis described above has been performed for all the $23$
samples obtained changing the film sizes, strains, compounds
($LSMO$ and $LCMO$) measuring $\rho$ in films fabricated with
several techniques or in films grown by the same technique with
different deposition parameters. In order to compare the different
results, we have chosen the same fitting temperature range
(20-100) K. Actually this temperature range is enough far from the
$MI$ transition temperature in order to avoid any spurious effect
due to vicinity of the $MI$ transition. At the same time, the $20$
K lower bound is quite large in order to avoid the effects in
temperature dependence of $\rho$ due to the upturn at low $T$. In
Fig. 2 we show the typical resistivity curves and relative fitting
values of some of the samples. There is evidence of a correlation
between the residual resistivity and fitting parameter $\alpha$.
Therefore the values of $\alpha$ versus $\rho_0$ are reported in
Fig. 3 for all the analyzed films. Quite surprisingly most films
present a value of $\alpha$ very close to $2.5$. In particular all
the samples in the range $0.04$ $m\Omega \cdot cm$ $<$ $\rho_0$
$<$ $1$ $m\Omega \cdot cm$ have a value equal to $2.5$ within the
estimated error bar. Our data show a deviation from $T^{2.5}$
dependence for both high and low $\rho_0$. In particular for  $1$
$m\Omega \cdot cm$ $<$ $\rho_0$ $<$  $10$ $m\Omega \cdot cm$  the
$\alpha$ exponent approaches the value $3$. Finally for $\rho_0$
$\le$ $0.4$ $m\Omega \cdot cm$ we find evidence of a tendency
towards small $\alpha$ values.

We notice that in the set of fabricated samples $\rho_0$ varies
from 0.03 $m\Omega \cdot cm$, that represents one of the lowest
values in manganites, to 6 $m\Omega \cdot cm$. Therefore all the
samples are characterized by a residual resistivity $\rho_0$
smaller than the Mott's maximum metallic resistivity that is of
the order of $\simeq$ 10 $m\Omega \cdot cm$ in these systems.
\cite{Tiwari} Moreover most films are characterized by
$\rho_0<\rho_c \simeq 1 m\Omega \cdot cm$, a critical value that
has been suggested to be a lower bound for the occurrence of an
Anderson $MI$ transition with increasing the temperature.
\cite{Sheng} As discussed above, for moderate disorder ($0.04$
$m\Omega \cdot cm$ $<$ $\rho_0$ $<$ $\rho_c$), the single-magnon
scattering assisted by the localized minority spin states explains
the transport properties in the low-$T$ range. However starting
just from $\rho_c$ a new scattering mechanism sets in. The value
$\alpha=3$ has been previously interpreted as due to an anomalous
single magnon scattering that can become dominant with the
decrease of spin-wave stiffness coefficient, that is proportional
to the one-electron bandwidth of the $e_g$ carriers.
\cite{Akimoto,Furukawa} With increasing the strength of the
disorder, it is possible that the effective bandwidth of the
itinerant charge carriers gets reduced. Therefore this new
transport regime is consistent with the increased strength of
disorder and is in agreement with previous experimental
investigations made on $Nd-$doped manganite systems with large
values of $\rho_0$. \cite{Furukawa1} Finally in the regime of
small disorder ($\rho_0 \simeq 0.03-0.04$  $m\Omega \cdot cm$) the
$\alpha$ exponent tends toward the value $2$ that is
characteristic of single-crystals.

In the next section we will analyze the transport properties at
high temperature pointing out the strong interplay between
disorder and $el-ph$ coupling in determining the insulating phase.
In fact it has been stressed that effects due disorder should not
be able alone to drive the $MI$ transition. \cite{Smolyaninova}
However, it has been also shown that effects due to disorder can
enhance the tendency toward the polaron formation and the
sensitivity to changes in the $el-ph$ coupling. \cite{Kumar1}

\subsection{Whole temperature range}

In this section we analyze the resistivity in the high and the
whole $T$ range.

Single crystals and optimized films of $LCMO$ show the $MI$
transition at close temperatures. Even if the strength of the
intrinsic disorder in these materials is not negligible, the
transport properties in the $PI$ phase are typically described in
terms of polaronic conduction stressing the role of the $el-ph$
interaction in driving the $MI$ transition. \cite{Salomon} For
$T>T_p$ the resistivity is characterized by an activated behavior
that can be described by the following law
\begin{equation}
\rho_{PI}(T)=\rho_{\infty} \cdot
exp\left(\frac{E_{g}}{K_{B}T}\right),
\label{eqma}
\end{equation}
with the activation energy $E_{g}$ of the order of $0.1-0.2$ $eV$.
In particular, in the insulating phase an high temperature
expansion of the polaronic resistivity gives the dependence
$\rho_{\infty} \propto \sqrt T$ and $E_{g}=E_{p}/2$, with $E_p$
polaron binding energy. \cite{perromanga} In the $PCS0$ sample the
best fit to $\rho$ is provided by the polaronic hopping mechanism.
Other forms such as those predicted by variable range hopping
\cite{Coey} were also used to fit the data, but they yield less
accurate fits ($R^2$ remarkably smaller than unity). In Fig. 4 the
plot of the resistivity is reported in the temperature range up to
$300$ $K$. For the $LCMO$ film the best fit is obtained for $E_g=
82.15$ $meV$ that is consistent with the results of previous
investigations. Clearly in this regime the role of the correlation
between polarons can be important since it gives rise to charge
ordering fluctuation. \cite{adams}

In order to interpret the transport properties in the intermediate
range of $T$, the effects of the phase separation between $FM$ and
$PI$ phases have been invoked. \cite{Salomon,Li,perrofilm} If the
properties of these systems are driven by the coexistence of $FM$
and $PI$ phases, \cite{perromanga} $\rho(T)$ can be written as
\begin{equation}
\rho(T)=\rho_{FM} \cdot\emph{f}+\rho_{PI}\cdot(1-\emph{f}),
\label{eqtot}
\end{equation}
where $\rho_{FM}$ is given by Eq. (\ref{exponent}) and $\rho_{PI}$
by the polaron hopping term of Eq. (\ref{eqma}). The function
\emph{f} represents the volume fraction of the $FM$ metallic
regions in the system while (1-\emph{f}) represents the
paramagnetic one. This function has a value equal to unity at low
temperatures, is decreasing with increasing $T$ and goes to zero
in the $PI$ phase. The fits of the data in the low and high
temperature region, given by Eq. (\ref{exponent}) and
(\ref{eqma}), respectively, are extended in the whole temperature
range, so we have extracted the distribution function \emph{f}
using for $\rho(T)$ in Eq. (\ref{eqtot}) the experimental data.
The function \emph{f} reported in the inset of Fig. 4 is in
qualitative agreement with the fraction of volume calculated
within a single-orbital model that takes into account the combined
effect of the magnetic and $el-ph$ interactions. \cite{perromanga}
Actually, in agreement with many other theoretical works and
experimental observations, \cite{dagotto} the combined effect of
these interactions pushes the system toward a regime of two
coexisting phases: one made by itinerant carriers forming
ferromagnetic domains and another by localized polarons giving
rise to paramagnetic or antiferromagnetic domains depending on
temperature.

The analysis of the transport properties in the whole temperature
range is challenging for $LSMO$ systems. In fact, single crystals
of $LSMO$ are metallic ($ d \rho /dT>0$) in any range of
temperature. However, even when the conduction is metallic, a high
temperature polaronic behavior is directly observed by means of
photoemission, x-ray absorption and emission, and extended x-ray
absorption fine structure spectroscopy. \cite{Mannella} Moreover,
unlike single crystals, in $LSMO$ films the $FP$ transition is
accompanied by a close $MI$ transition, as in $LCMO$ systems. The
role played by both $el-ph$ coupling and disorder can be crucial
in stabilizing the insulating phase. Actually it has been shown
that there is a positive feedback of disorder on the polaron
formation and an increase of the sensitivity of the system to
variations of $el-ph$ coupling. \cite{Kumar2} In a regime of
moderate disorder, at high temperatures the system can change from
a metallic ($d\rho/dT>0$) to an insulating ($d\rho/dT<0$) behavior
by means of a slight increase of the $el-ph$ coupling. Therefore
in the films, where effects of disorder can be stronger and the
strain is able to increase the $el-ph$ coupling,
\cite{Chen,perrofilm} the interplay of disorder and $el-ph$
coupling can be able to drive a $MI$ transition absent in the
single crystal bulk case.

The $LSMO$ films grown by different techniques and analyzed in
this work show $MI$ transition temperatures ranging from $300$ $K$
to values slightly larger than $400$ $K$. We find that the value
of $T_p$ generally increases as the residual resistivity decreases
even if this $MI$ transition temperature is strongly dependent
also by other factors such as the film thickness, the orientation,
and the value of the strain. Therefore it not easy to recognize a
clear relation between $T_p$ and $\rho_0$ unless the other
parameters are under strict control. In Fig. 5 we show the results
obtained on two different doped samples: $SSS0$ grown on (100)
$STO$ and $SSS1$ on (110) $STO$. These two samples have residual
resistivities smaller than $\rho_c \simeq 1 m\Omega \cdot cm$,
therefore at low temperature the temperature dependence of $\rho$
is dominated by the $T^{2.5}$ contribution. At high $T$ both
$SSS0$ and $SSS1$ resistivities show an activated behavior, so the
best two-parameter fit is given by Eq. (\ref{eqma}). Moreover we
have found that the parameters considered in variable range
hopping mechanism, such as the localization length, show
variations of many orders of magnitude for films with close values
of residual resistivity and critical temperature (for example data
shown in Fig. 5). The sample $SSS1$ shows a sharp maximum in the
resistivity that in the range $500-800$ $K$ is well described by
Eq. (\ref{eqma}) with $\rho_{\infty} \propto {\sqrt T }$ and an
activated energy $E_g$ equal to $64.37$ $meV$. Instead the sample
$SSS0$ is on the verge of the metallic phase, in fact the
resistivity is weakly decreasing and the activation energy is an
order of magnitude smaller than that of $SSS1$. Therefore the
different behavior of the resistivities of two samples correlates
with the decrease of the residual resistivity. Indeed for $LSMO$
films with $\rho_0$ smaller than $0.1$ $ m\Omega \cdot cm$ and
$T_p$ larger than $400$ $K$ the resistivity is characterized by a
broad maximum around $T_p$ and it decreases very slowly as
function of the temperature in the insulating side. Finally, on
the basis of recent investigations reporting phase separation also
in $LSMO$ films, \cite{Mannella,Hartinger} we propose to interpret
the resistivity data on the whole temperature range employing Eq.
(\ref{eqtot}). Following the same procedure used for $LCMO$ films,
we can extract the distribution function \emph{f} that provides
the volume fraction of the $FM$ phase in the system. The
distribution functions for the two samples (inset of Fig. 5) bear
a strong resemblance with those obtained in the case of the $LCMO$
films, in fact there is only a slower variation in temperature.
Hence these data seem to confirm that the phase separation
scenario can adopted also in the analysis of the transport
properties of $LSMO$ films.

Comparing Fig. 4 e Fig. 5, there is also a correlation between the
residual resistivity and the activation energy. By increasing
$\rho_0$, the samples are characterized by a larger activation
energy that is a measure of the coupling of the charge carriers to
the lattice. Therefore these data confirm the interplay of
disorder and $el-ph$ coupling that represent key parameters in
order to understand the properties of these materials and in
particular the $CMR$ effect.
\cite{dagotto,Akahoshi,Motome,dagotto1}

\section{Summary}
In this paper we have discussed the transport properties of $LCMO$
and $LSMO$ films for temperatures up to $800$ $K$. We have made
the comparison of the results between films grown by different
techniques since this gives the possibility to investigate samples
with different amounts of disorder remaining in the $FM$ phase.
The first part of our analysis has focused on the low temperature
range where we have found clear evidence that the temperature
contribution scales as $T^{\alpha}$ with $\alpha$ close to $2.5$
for an intermediate range of residual resistivities. For large
values of disorder the temperature dependence of the resistivity
fits well the law $T^{3}$ characteristic of anomalous single
magnon scattering processes, while in the regime of low strength
of disorder $\alpha$ shows a tendency towards a value near $2$.
These results is independent of the film thickness, on the strain
distribution and on the growth technique, and supports the role of
the single magnon scattering. At high temperatures the activated
behavior of polaronic carriers represents the prominent behavior
in most films where the disorder seems to increase the tendency
toward the polaron formation and correlates with the activation
energy. In the whole range of temperatures the experimental data
seem to support a phase separation scenario that has been proposed
by recent studies also in $LSMO$ systems.

In order to further elucidate the low $T$ behavior of resistivity,
it would be interesting to pursue the study of the transport
properties in presence of magnetic field. When an external field
is applied, the disorder is expected to be reduced influencing not
only the upturn around $10-20$ $K$ \cite{Kumar,Ziese} but also the
single magnon scattering. \cite{Wang} The analysis in magnetic
field will be the subject of a future study.

\section{Acknowledgments}
This work has been partially supported by PRIN project "Strain
effects on the metal-insulator transition and on the metallic
state of manganite thin films and heterostructures" of the Italian
Minister of the University and Research (MIUR). S. Mercone, C.
Adamo, and A. Oropallo are supported by "Regione Campania" within
the project "Creazione di operatori per il trasferimento
tecnologico da enti pubblici di ricerca a piccole e medie imprese"
managed by CRdC ('Centro Regionale di Competenza') of Universita'
di  Napoli "Federico II". P. Perna and U. Scotti di Uccio
acknowledge partial support by the MIUR under the PRIN project
"Sviluppo di materiali avanzati e nuove tecnologie di produzione
per applicazioni nel campo della sensoristica".

\section*{Figure captions}
\begin{description}

\item{F1}
Low-temperature resistivity in three different  ranges: (a) range
between 4 $K$ and 60 $K$, (b) range between 10 $K$ and 120 $K$, and
(c) range between 10 $K$ and 200 $K$. Experimental data of $PSS0$
(open circle) and $MSS0$ (open triangle) films are shown with the
corresponding fits (line).

\item{F2}
Low-temperature resistivities with corresponding fits in the range
between 20K and 100K for different films. The fits are obtained by
equation (\ref{exponent}) where the fit exponent $\alpha$ is
defined.

\item{F3}
Low-temperature fit exponent $\alpha$ as function of the residual
resistivity $\rho_0$.

\item{F4}
Resistivity of the $PCS0$ film (solid line) as function of the
temperature. The dotted line stands for the low-temperature fit,
while the dashed line for the high-temperature one. In the inset the
distribution function \emph{f} derived through equation
(\ref{eqtot}) is reported.

\item{F5}
Resistivities of the $SSS0$ and  $SSS1$ as function of the
temperature. The dotted lines stand for the low-temperature fits,
while the dashed line for the high-temperature ones. In the inset
the distribution functions \emph{f} derived through equation
(\ref{eqtot}) for $SSS0$ (dot-dash line) and $SSS1$ (solid line)
films are plotted.

\end{description}

\newpage

\begin{table*}[t]
\centering
  \begin{tabular}[t]{c|c|c|c|c|c|c|c}
  \hline\hline
  \textsc{Technique} & \textsc{Name} & \textsc{Composition} & \textsc{Substrate\textnormal{(h,k,l)}} & \textsc{Thickness(\AA)} & \textsc{$T_{p}(K)$}
  & \textnormal{c axis (\AA)} & Rocking width \\
  \hline \hline
  PLD & {\bf PCS0} & $La_{0.7}Ca_{0.3}MnO_{3}$ & $SrTiO_{3}$(100) & 3000 & $(245\pm1)$ & $3.86\pm0.01$ & $0.2^o$ \\
  \hline
  PLD & {\bf PSL0} & $La_{0.7}Sr_{0.3}MnO_{3}$ & $LaAlO_{3}$(100) & 1600 & $(364\pm2)$ & $3.87\pm0.01$ & $0.2^o$ \\
  \hline
  MBE & {\bf MSS0} & $La_{0.7}Sr_{0.3}MnO_{3}$ & $SrTiO_{3}$(100) & 350 & $(344.6\pm0.1)$ & $3.79\pm0.01$ & $0.1^o$  \\
  \hline
  MBE & {\bf MSN1} & $La_{0.7}Sr_{0.3}MnO_{3}$ & $NdGaO_{3}$(110) & $210$ & $(>400)$ & $3.9\pm0.1$ & $ 0.1^o$
   \\
  \hline
  Sputtering & {\bf SSS1} & $La_{0.7}Sr_{0.3}MnO_{3}$ & $SrTiO_{3}$(110)& 400 & $(400\pm1)$ & ****       & $ <0.1^o$\\
   \hline
  Sputtering & {\bf SSS0}   & $La_{0.7}Sr_{0.3}MnO_{3}$   & $SrTiO_{3}$(100)& 400 & $(350\pm1)$ & $3.85\pm0.01$ & $ <0.1^o$\\
   \hline
  Sputtering & {\bf SSL1} & $La_{0.7}Sr_{0.3}MnO_{3}$ & $LaAlO_{3}$(110) & 400 & $(380\pm1)$ & **** & $ <0.1^o$\\
  \hline\hline
\end{tabular}
\caption{Representative samples obtained by different fabrication
techniques. For the films with (110) orientation we have reported
the values of the lattice parameters in the text.} \label{samples}
    \end{table*}

\begin{table}[t]
\centering
  \begin{tabular}[t]{|c|c|c|c|c|c|c|}
  \hline
  Sample & Range T (K) & {\bf $\rho_{0}(\Omega*cm)$} & {\bf A$(\Omega*cm*K^{-\alpha})$} & {\bf $\alpha$}  & {\bf $R^2$}\\
  \hline
  PCS0 & $4-60$  & $0.00103$ & $4.33*10^{-9}$ & $2.58$ & 0.99892\\
  PCS0 & $4-120$ & $0.00103$ & $5.31*10^{-9}$ & $2.53$ & 0.99970\\
  PCS0 & $4-150$ & $0.00104$ & $2.49*10^{-9}$ & $2.69$ & 0.99960\\
  MSS0 & $20-60$ & $0.00244$ & $4.47*10^{-9}$ & $2.69$ & 0.99905\\
  MSS0 & $20-120$& $0.00243$ & $1.16*10^{-8}$ & $2.48$ & 0.99953\\
  MSS0 & $20-200$& $0.00244$ & $8.38*10^{-9}$ & $2.55$ & 0.99973\\
  \hline
\end{tabular}
\caption{Samples obtained from different growth techniques
analyzed in a wide range of temperatures.} \label{fittable}
 \end{table}

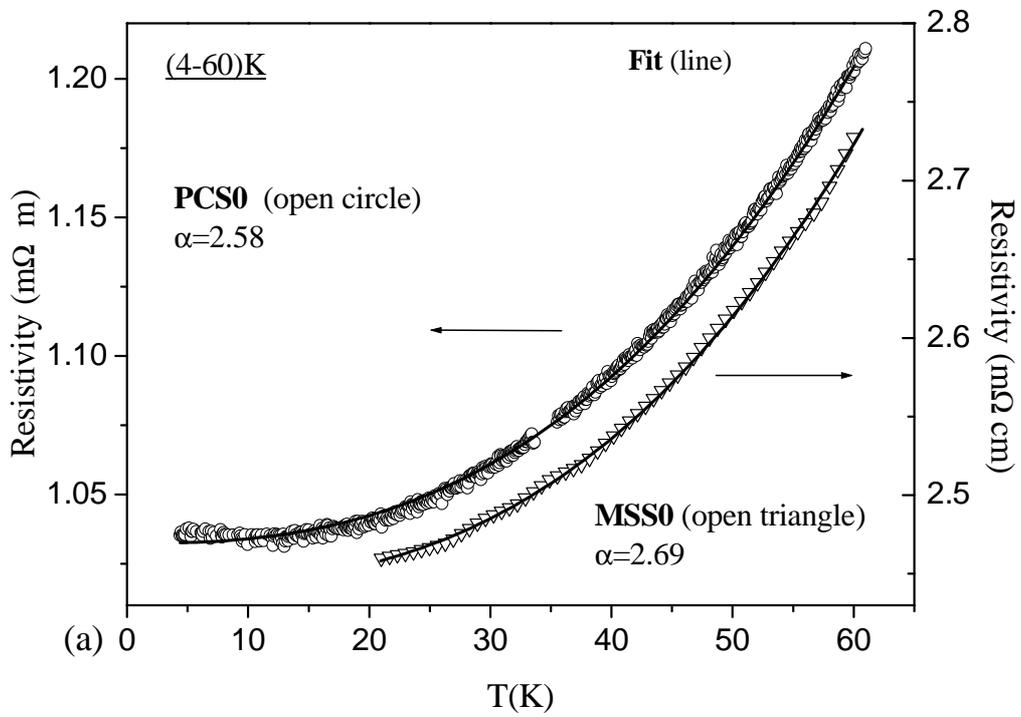
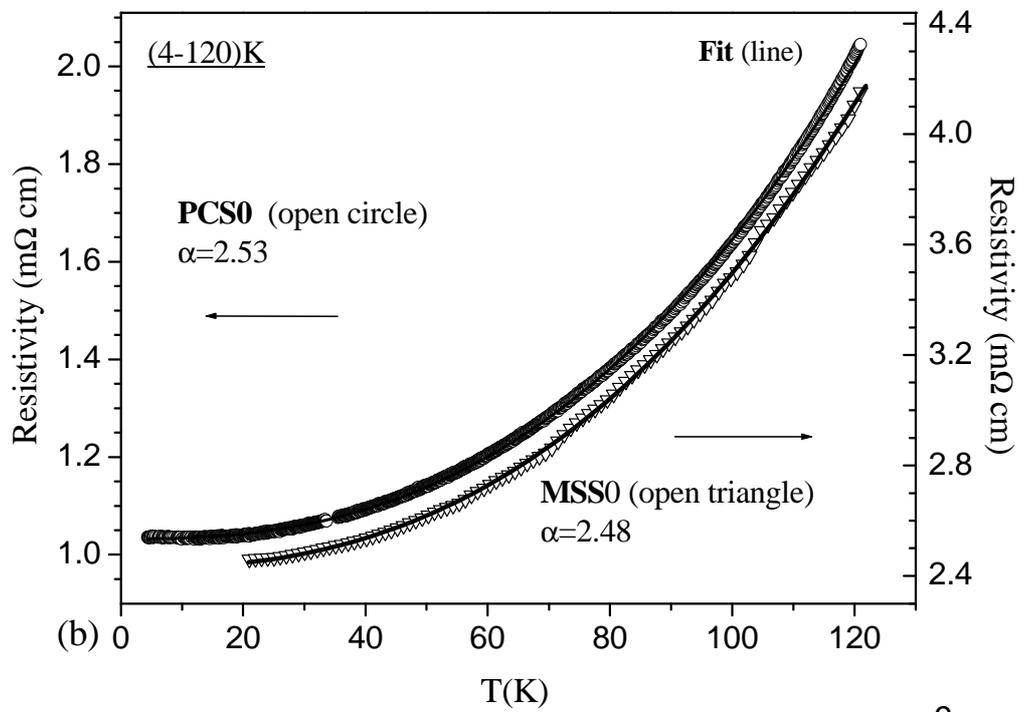
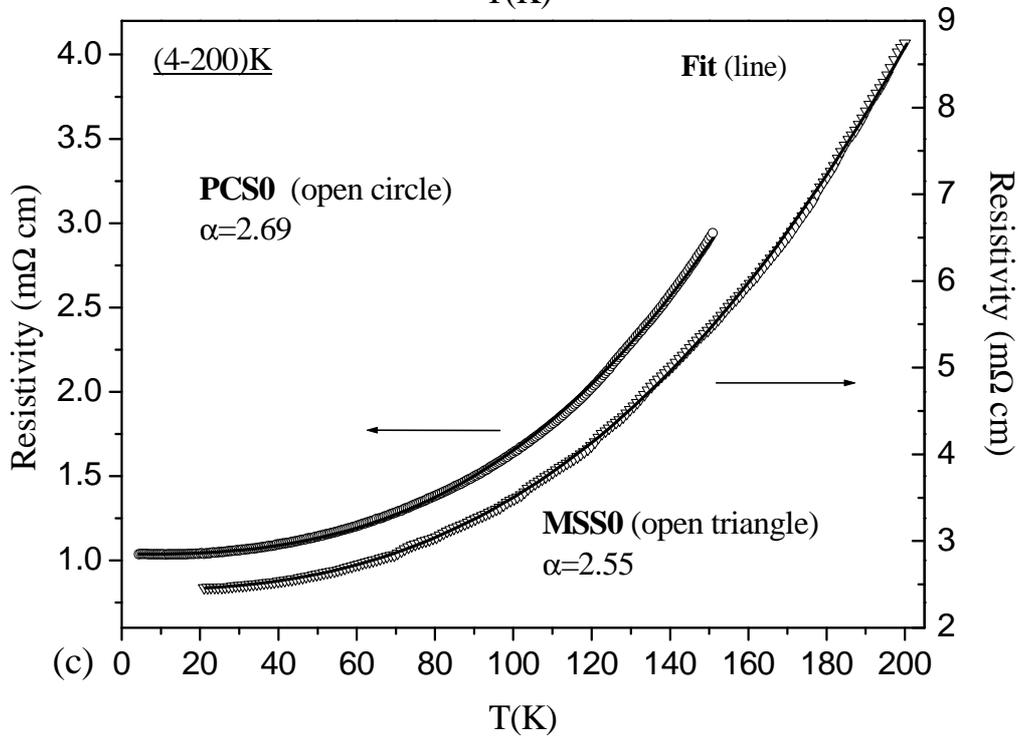
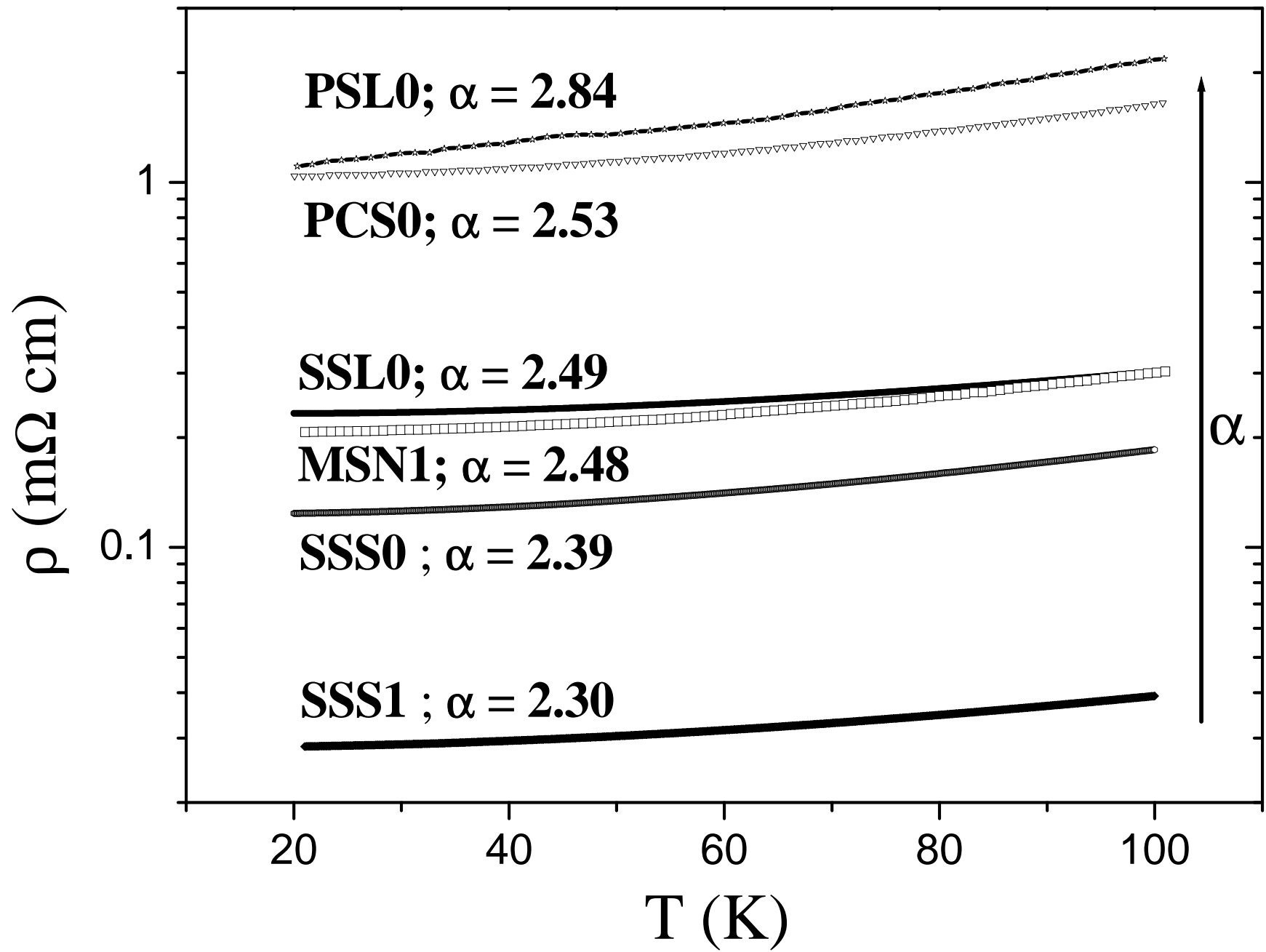
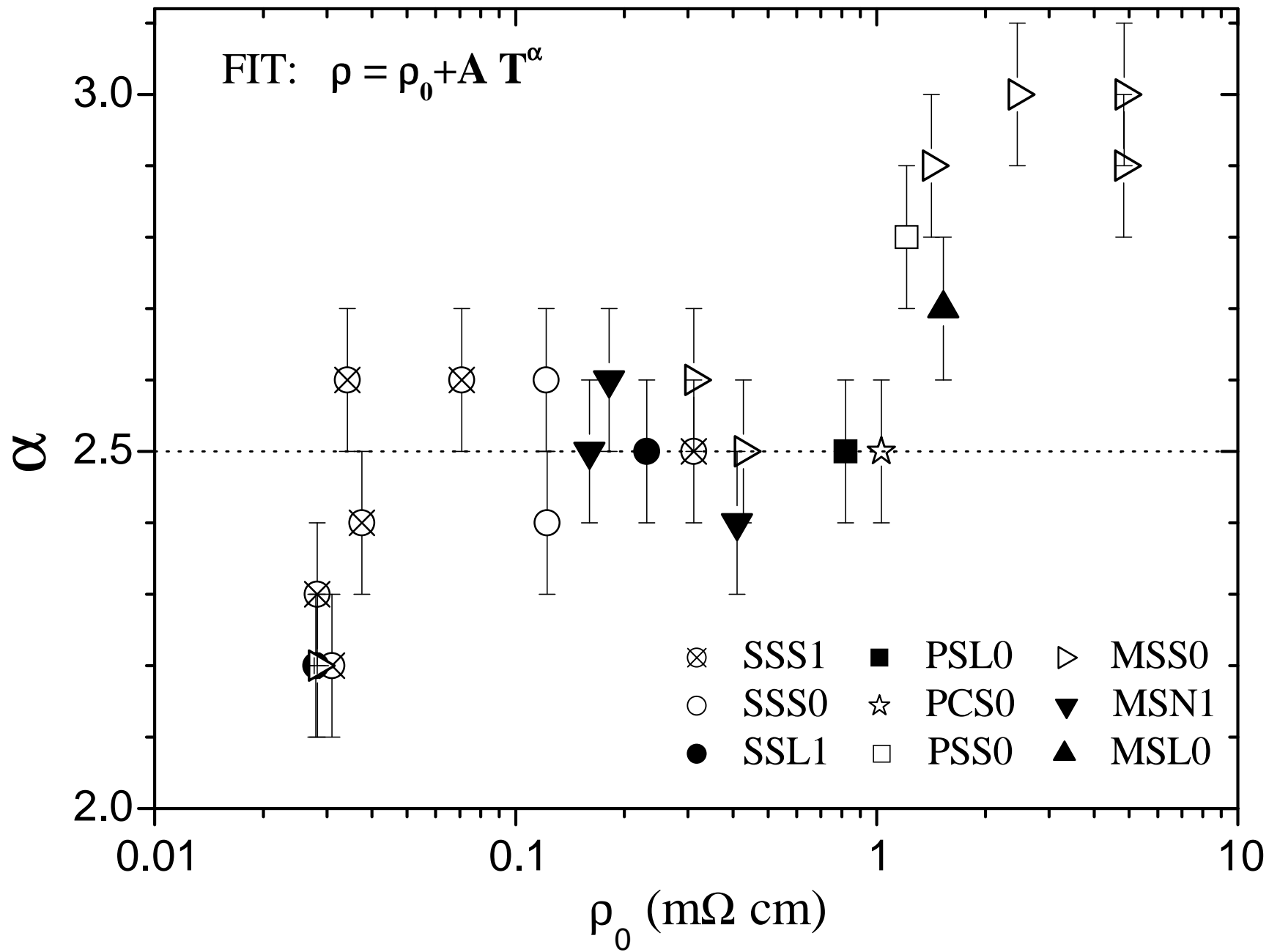
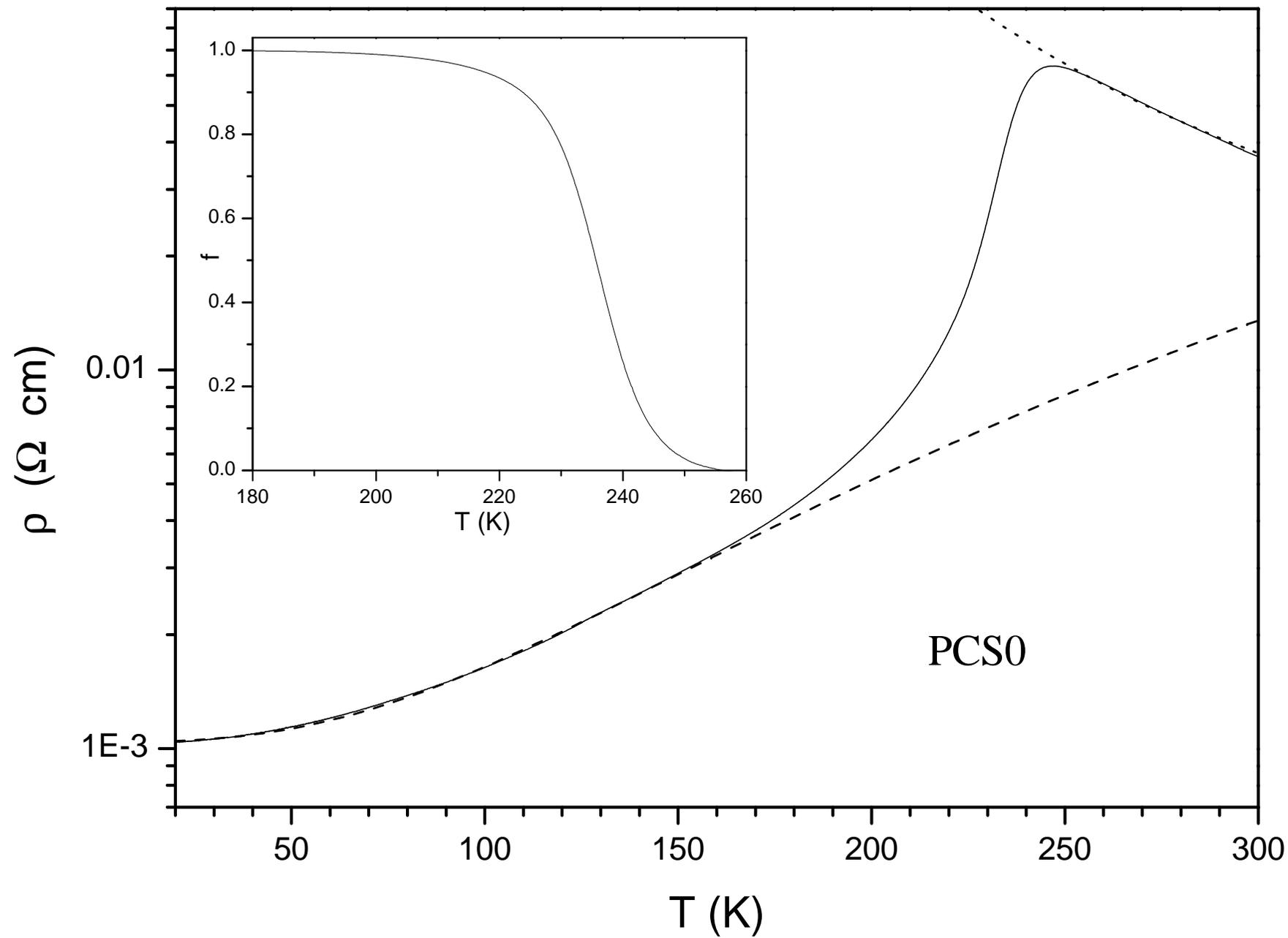
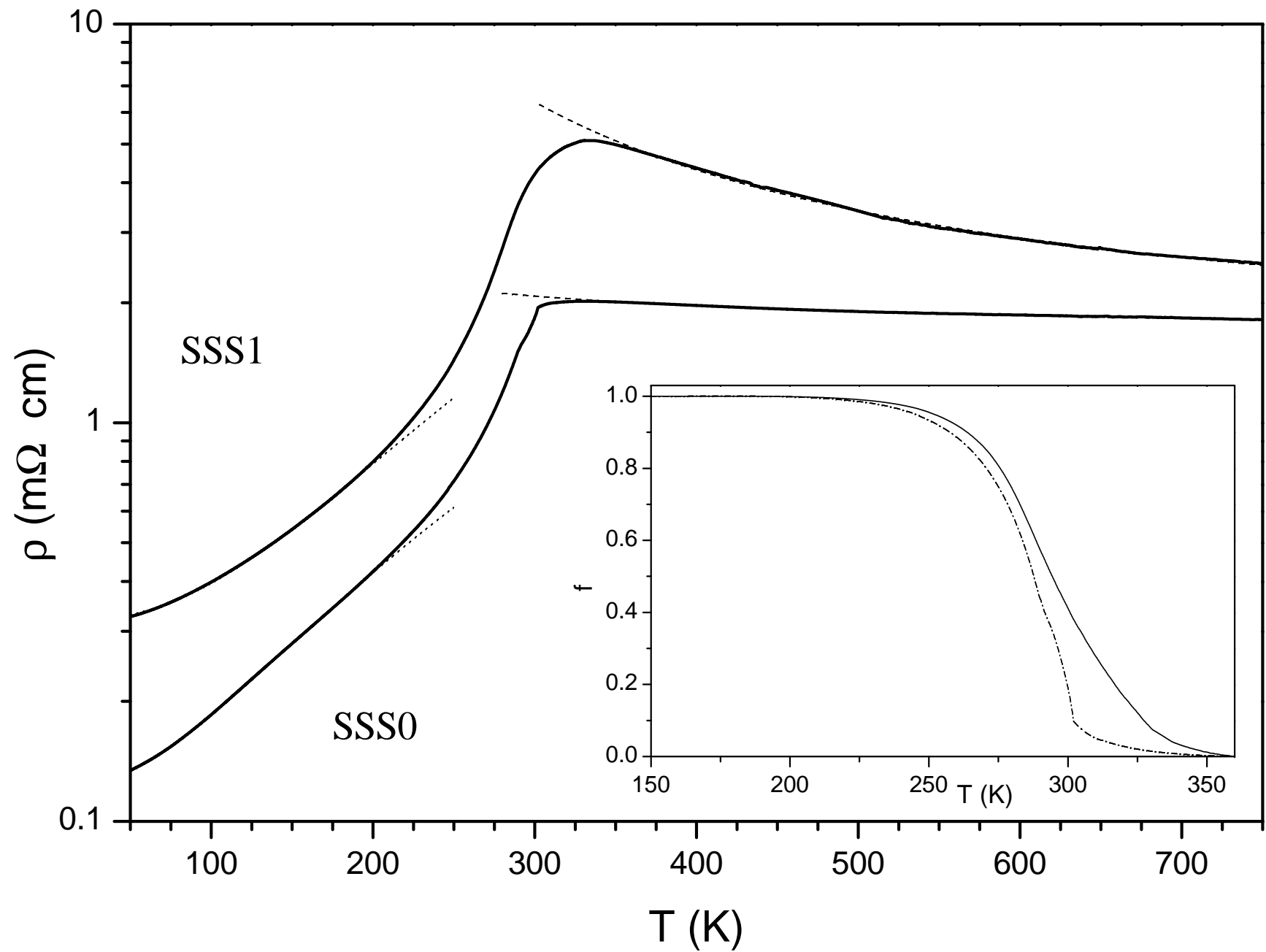
\end{document}